\documentstyle{comments}

\newif\ifAMStwofonts

\def\fun#1#2{\lower3.6pt\vbox{\baselineskip0pt\lineskip.9pt
  \ialign{$\mathsurround=0pt#1\hfil##\hfil$\crcr#2\crcr\sim\crcr}}}

\def\rati{M_h/M_g}
\def\msun{{M$_\odot$\ }}

\ifoldfss
  \ifCUPmtlplainloaded \else
    \NewTextAlphabet{textbfit} {cmbxti10} {}
    \NewTextAlphabet{textbfss} {cmssbx10} {}
    \NewMathAlphabet{mathbfit} {cmbxti10} {} 
    \NewMathAlphabet{mathbfss} {cmssbx10} {} 
  \fi
  \ifAMStwofonts
    \ifCUPmtlplainloaded \else
      \NewSymbolFont{upmath} {eurm10}
      \NewSymbolFont{AMSa} {msam10}
      \NewMathSymbol{\upi}     {0}{upmath}{19}
      \NewMathSymbol{\umu}     {0}{upmath}{16}
      \NewMathSymbol{\upartial}{0}{upmath}{40}
      \NewMathSymbol{\leqslant}{3}{AMSa}{36}
      \NewMathSymbol{\geqslant}{3}{AMSa}{3E}

    \fi
  \fi
\fi 

\ifnfssone
  \newmathalphabet{\mathit}
  \addtoversion{normal}{\mathit}{cmr}{m}{it}
  \addtoversion{bold}{\mathit}{cmr}{bx}{it}
  \newmathalphabet{\mathbfit} 
  \addtoversion{normal}{\mathbfit}{cmr}{bx}{it}
  \addtoversion{bold}{\mathbfit}{cmr}{bx}{it}
  \newmathalphabet{\mathbfss} 
  \addtoversion{normal}{\mathbfss}{cmss}{bx}{n}
  \addtoversion{bold}{\mathbfss}{cmss}{bx}{n}
  \ifAMStwofonts
    \ifCUPmtlplainloaded \else
      \UseAMStwoboldmath
      \makeatletter
      \new@mathgroup\upmath@group
      \define@mathgroup\mv@normal\upmath@group{eur}{m}{n}
      \define@mathgroup\mv@bold\upmath@group{eur}{b}{n}
      \edef\UPM{\hexnumber\upmath@group}
      \new@mathgroup\amsa@group
      \define@mathgroup\mv@normal\amsa@group{msa}{m}{n}
      \define@mathgroup\mv@bold\amsa@group{msa}{m}{n}
      \edef\AMSa{\hexnumber\amsa@group}
      \makeatother
      \mathchardef\upi="0\UPM19
      \mathchardef\umu="0\UPM16
      \mathchardef\upartial="0\UPM40
      \mathchardef\leqslant="3\AMSa36
      \mathchardef\geqslant="3\AMSa3E
    \fi
  \fi
\fi 

\ifnfsstwo
  \DeclareMathAlphabet{\mathbfit}{OT1}{cmr}{bx}{it}
  \SetMathAlphabet\mathbfit{bold}{OT1}{cmr}{bx}{it}
  \DeclareMathAlphabet{\mathbfss}{OT1}{cmss}{bx}{n}
  \SetMathAlphabet\mathbfss{bold}{OT1}{cmss}{bx}{n}
  \ifAMStwofonts
    \ifCUPmtlplainloaded \else
      \DeclareSymbolFont{UPM}{U}{eur}{m}{n}
      \SetSymbolFont{UPM}{bold}{U}{eur}{b}{n}
      \DeclareSymbolFont{AMSa}{U}{msa}{m}{n}
      \DeclareMathSymbol{\upi}{0}{UPM}{"19}
      \DeclareMathSymbol{\umu}{0}{UPM}{"16}
      \DeclareMathSymbol{\upartial}{0}{UPM}{"40}
      \DeclareMathSymbol{\leqslant}{3}{AMSa}{"36}
      \DeclareMathSymbol{\geqslant}{3}{AMSa}{"3E}
    \fi
  \fi
\fi 

\ifCUPmtlplainloaded \else
  \ifAMStwofonts \else 
    \def\upi{\pi}
    \def\umu{\mu}
    \def\upartial{\partial}
  \fi
\fi

\title{Galaxies and Black Holes}
\author[David Merritt]
       {David Merritt \\
Department of Physics and Astronomy, Rutgers University,
New Brunswick, NJ 08855.}
\date{Rutgers Astrophysics Preprint Series No. 231}

\pubyear{1997}

\begin{document}

\maketitle

\label{firstpage}


The idea that supermassive black holes are generic components of
galactic nuclei has come to be widely accepted, due largely to 
the secure detection of $10^6-10^{9.5}$\msun dark objects at the 
centers of about a dozen galaxies.$^1$
The mean mass of these objects -- of order $10^{-2.5}$ times 
the mass of their host galaxies -- is consistent with the 
mass in black holes needed to produce the observed energy density in 
quasar light given reasonable assumptions about the efficiency of 
quasar energy production.$^{2,3}$
The black hole paradigm also explains in a natural way many of the 
observed properties of energy generation in active galactic 
nuclei (AGNs),$^4$ 
and a model in which elliptical galaxies form from the mergers of 
disk galaxies whose bulges contain black holes has been shown to be 
consistent with the so-called ``core fundamental plane,''
the relation between the central parameters of early-type galaxies.$^5$
However it has long been clear that the dynamical influence of a 
supermassive black hole can extend far beyond the nucleus
if a substantial number of stars are on orbits that carry them into 
center.$^{6,7}$
Recent work, discussed here, has expanded on this idea and given support
to the view that nuclear black holes may be important for understanding 
many of the systematic, large-scale properties of galaxies, including 
the absence of bars in most disk galaxies, the shapes of spiral galaxy 
rotation curves, and the fact that elliptical galaxies come in two, 
morphologically-distinct families.
To the extent that the growth of black holes is dependent on 
the global morphology of their host galaxies, this link between
black holes and galaxy structure may imply a feedback 
mechanism that determines what fraction of a galaxy's mass ends up
in the central singularity.

\begin{figure*}
\vspace{10cm}
\includegraphics{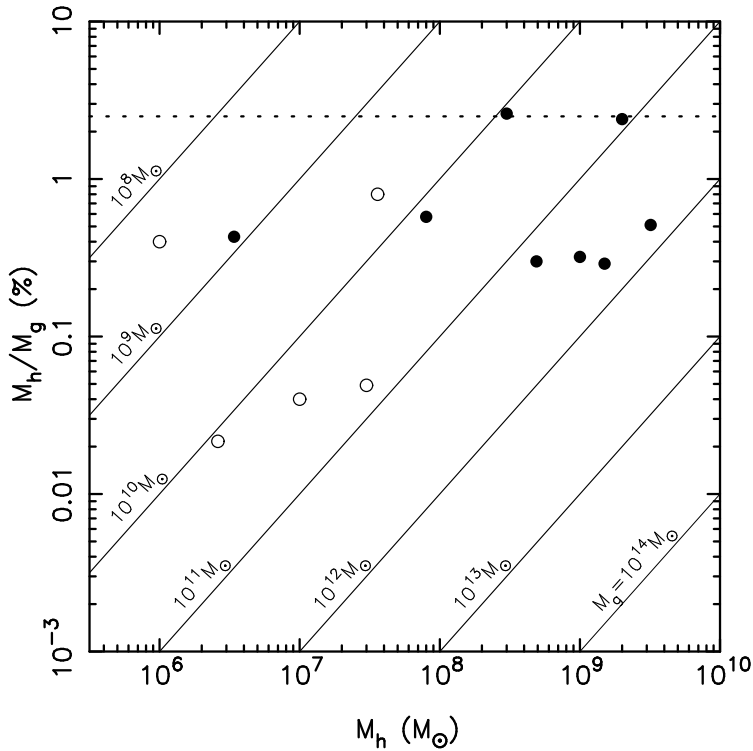}
\caption[]
{Black hole masses $M_h$, in units of the solar mass, and mass 
fractions $\rati$, with $M_g$ the mass of the bulge, for 13 
galaxies with securely-detected black holes.$^9$
Filled circles are early-type galaxies (S0, E, Sa); 
open circles are late-type galaxies (Sb, Sc).
The dashed line is the critical mass ratio that enforces
axisymmetry in a triaxial bulge.$^{35}$}
\label{fig1}
\end{figure*}

Dressler \& Richstone$^8$ pointed out a decade ago that the ratio 
of the masses of the dark objects in M31 and M32 was $5-10$, ``closer to 
the ratio of spheroid luminosities ($\sim15$) than it is to the ratio 
of total luminosities ($\sim70$).''
Since then, the approximate proportionality between black hole 
mass and bulge mass has held up fairly well. 
Figure 1, based on Richstone's$^9$ recent compilation, 
shows black hole masses $M_h$ vs. mass ratios $M_h/M_g$ 
for 13 galaxies; $M_g$ is defined as the 
total luminous mass in the case of elliptical galaxies and as the 
bulge mass in the case of disk galaxies.
(Two galaxies with questionable black
hole detections, NGC 3379 and NGC 4486b, were omitted.)
Figure 1 suggests a typical value for $\rati$ of $\sim0.5\%$ and an 
upper limit of $\sim3\%$.
There is a trend of increasing $\rati$ with $M_h$ which probably
just reflects the shortage of very bright and very faint galaxies 
in the sample.
Perhaps more significant is the tendency of $\rati$ to depend 
on Hubble type: the largest mass fractions are seen in early-type galaxies, 
S0's and E's, and the smallest in late-type galaxies, Sb's and Sc's.
This trend is reproduced even within the (essentially complete)
sample of galaxies from the Local Group:
$\rati$ for the Milky Way and M31 are $0.02\%$ and $0.05\%$
while the elliptical galaxy M32 has $\rati\approx0.4\%$.
(M33 has neither a detectable black hole nor a detectable bulge$^{10}$).

Uncertainties in the determination of $M_h$ are often large$^1$
but bulge masses can be very uncertain too.
One source of systematic error that might introduce a spurious 
dependence of $\rati$ on Hubble type has been pointed out by 
Kormendy.$^{11}$
Bulges are difficult to disentangle photometrically from disks and 
the decomposition is most difficult in late-type spirals which have 
the smallest bulge-to-disk ratios.
A blind decomposition assuming an exponential disk tends to give 
spuriously large bulge luminosities, hence overly small values of 
$\rati$, since many disks have steeper-than-exponential profiles 
near the center.
The luminosity and mass of the Milky Way bulge, which has the 
lowest value of $\rati$ in Figure 1, are especially poorly 
known.
The bulge mass used in Figure 1 was taken from Kent's$^{12}$ model
which was based on $K$-band photometry with (large) corrections for 
absorption and for the light from the disk.
While Kent's model accounts nicely for the kinematics of a wide range 
of bulge tracers, some components of the bulge are observed to fall off 
more steeply$^{13,14}$ suggesting a 
possibly smaller total mass and a larger $\rati$.

Nevertheless it is reasonable to expect lower values of 
$\rati$ in galaxies with small bulge-to-disk ratios since 
these galaxies are the least likely to have experienced the 
strong non-axisymmetric distortions that are believed necessary 
for driving large amounts of gas into the nucleus and feeding the
black hole.$^{15}$
For instance, if tidal encounters or mergers during the quasar epoch 
were responsible for black hole growth,$^{16}$
the smallest mass accumulations would be expected in late-type spirals 
which have presumably suffered the fewest such interactions.
Other factors that might limit the accumulation of mass 
in the nuclei of late-type galaxies are their generally weaker 
bars$^{17}$ and the fact that gas flows in barred 
galaxies with small bulges tend not to form shocks.$^{18}$

The argument that large-scale nonaxisymmetries in the stellar 
distribution are necessary for the formation of black holes 
is based on the stringent requirements that 
quasar luminosities place on fueling mechanisms.$^{15}$
Gravitational torques from a stellar bar -- induced by galaxy 
interactions or internal self-gravitational instabilities, or 
both -- can remove much of the angular momentum from the 
interstellar medium on kiloparsec scales, allowing a large 
fraction of a galaxy's gas to approach the nucleus in a dynamical 
time.
This picture derives some support from the observation that the 
host galaxies of AGNs often have close companions or substantial
asymmetries.$^{19,20}$
However the evidence for a direct link between stellar bars and 
nuclear activity is weak at best.
Neither the detection rate nor the emission-line strength of AGNs 
appear to be influenced by the presence of a stellar bar,$^{21,22}$
and the majority of barred galaxies exhibit no significant nuclear 
activity.

The relevance of such data to theories of black hole formation
is difficult to judge because the bars that we see now need not
be the ones that were responsible for creating the black holes.
Indeed, the accumulation of mass at the center of a barred galaxy 
can weaken or even dissolve the bar by changing the major families 
of orbits.$^{23,24}$
$N$-body simulations$^{25,26}$ suggest that bar disruption 
follows rapidly after several percent of a disk's mass has 
accumulated in a compact central object.
This mechanism may help to account for the scarcity of 
luminous AGNs in the current universe$^{27}$ but
it does not shed light on why barred and 
unbarred galaxies show roughly the same level of activity.

Sellwood \& Moore$^{28}$ have recently proposed a model that 
reconciles the bar-driven formation of black holes at early
times with the apparent inability of bars to generate nuclear 
activity at the current epoch.
They simulated the evolution of a stellar disk in which a central 
mass was grown at a rate that depended on the strength of the 
bar.
They found, in agreement with earlier studies, that the bar 
was substantially weakened once a few percent of the disk mass 
accumulated at the center.
As the central object grew, it pulled in matter from the disk, 
forming a bulge with much greater total mass.
Sellwood \& Moore then simulated infall of matter from the 
halo onto the disk by adding additional particles on circular 
orbits at large radii.
Even after the mass of the disk had grown by a factor of 
$\sim4$, no bar formed because the inner Lindblad resonance resulting 
from the newly-formed bulge suppressed the bar-making instability.$^{29}$
The galaxy that resulted from this experiment resembled a 
late-type, unbarred spiral with a realistically flat inner
rotation curve.
Sellwood \& Moore suggest that bars may nevertheless 
still form in real galaxies due to large perturbations or tidal 
encounters that can overwhelm the damping effect of the resonance.
Such bars would be prohibited by the ILR from channeling gas 
into the nucleus, however, thus limiting their activity in the current 
universe.

Large-scale stellar bars like those in Sellwood \& Moore's 
simulations can reduce the angular momentum of the gas 
by only an order of magnitude or so.$^{30}$
How the gas is funneled into the inner few parsecs is unclear 
but such transport probably requires non-axisymmetric distortions 
of the stars and gas on still smaller scales.$^{31-33}$
Any mechanism that enforced axisymmetry in the dominant mass
component of the bulge would 
therefore be expected to limit the mass of a central object 
regardless of the presence of a large-scale bar.
One such mechanism has long been known. 
The box orbits that support a triaxial stellar system are strongly 
influenced by a central point mass, since stars on box orbits pass 
arbitrarily close to the center after a long enough time.$^{6,7}$ 
A triaxial bulge containing a central black hole would be 
expected to evolve toward rounder or more axisymmetric shapes 
as the box orbits gradually lost their distinguishability, in 
much the same way that bars are weakened or destroyed by
central mass concentrations.

An early study$^{34}$ of the effect of a central point mass on a triaxial 
galaxy found only a modest degree of evolution,
but the softening length used was comparable to the radius 
of influence of the black hole and so the degree of orbital evolution 
was probably strongly underestimated.
Much more dramatic evolution was seen in a recent study$^{35}$
based on a hybrid $N$-body code 
with individual particle timesteps.
The authors found rapid evolution to nearly complete axisymmetry 
once the central mass exceeded $\sim 2.5\%$ the mass of the galaxy.
This mass ratio is interestingly close to the largest values observed 
in real galaxies (Figure 1): the current record-holders, 
NGC 3115 and NGC 4342, both have $\rati\approx 0.025$.
The agreement suggests that $\rati$ may be limited by a feedback 
mechanism that turns off the supply of fuel when the black hole grows 
large enough to force its bulge into an axisymmetric shape.
Stated differently, black holes may sometimes grow as 
large as the morphology of their host spheroids permit.

It is clear from Figure 1 that not all bulges contain black holes 
with $M_h\approx 0.025M_g$, but it is not difficult to think of 
additional mechanisms that would limit or reduce $\rati$.
The efficiency with which stellar bars drive gas into the bulge 
could differ from galaxy to galaxy, as discussed above.
The central parts of some galaxies might reach an axisymmetric 
state through processes unrelated to the presence of a black 
hole.$^{36}$
On smaller scales, the back-reaction of the energy emitted from 
the black hole on the accretion flow might limit black hole 
masses.$^{37,38}$
Successive mergers between galaxies already containing 
supermassive black holes could result in a gravitational 
slingshot if a third black hole is brought into a nucleus that 
contains an uncoalesced binary.$^{39}$
Mergers after the quasar epoch would also tend to reduce the 
average value of $\rati$ by converting disks to spheroids.
Typical disk-to-bulge ratios for S0 galaxies like NGC 3115 
and NGC 4342 are of order unity;
a merger between two such galaxies would reduce $M_h/M_g$ by 
a factor of $\sim2$ (assuming no further growth of the black 
holes), perhaps accounting for some of the vertical scatter in Figure 1.

Dissipationless mergers between disk galaxies tend to create 
strongly triaxial objects.$^{40}$
Adding a dissipative component greatly reduces the triaxiality,$^{41,42}$ 
presumably via destabilization of the box orbits as gas deepens the potential.
Similar behavior would be expected following the merger 
of two galaxies containing black holes: the merger-induced triaxiality
would gradually be destroyed by the central black hole at a rate 
determined by $\rati$ in the merged system.
Merritt \& Quinlan$^{35}$ find that a black hole with $\rati=0.3\%$, 
close to the average value for the elliptical galaxies in Figure 1, 
destroys triaxiality in roughly $100$ periods of the half-mass 
orbit.
This works out to $\sim5\times 10^9$ years for a typical 
early-type galaxy with $M_B=-19$ using observed scaling 
relations.$^{43-45}$
Fainter galaxies are denser, i.e. have shorter crossing times.
It follows that most elliptical galaxies fainter than $M_B\approx -19$ 
should have evolved to axisymmetric shapes by now (even if they 
were not axisymmetric initially), 
while brighter ellipticals could still be triaxial.
In fact the Hubble-type distribution
of elliptical galaxies undergoes a systematic change at about this 
luminosity and bright ellipticals appear to be moderately triaxial as a 
class.$^{46}$
The sudden increase of elliptical galaxy radio luminosity at 
$M_B\approx -19$ has also been attributed to a 
greater degree of triaxiality among bright ellipticals.$^{47,48}$ 

Bright and faint ellipticals differ also in the steepness of
their central luminosity profiles: faint ellipticals have $\rho\sim 
r^{-2}$ near the center while bright ellipticals have 
shallower cusps, $\rho\sim r^{-1}$ -- $r^0$.$^{49-51}$ 
Density cusps would themselves induce chaos in the box orbits
of a triaxial galaxy,$^{52,53}$
and the steep cusps in faint ellipticals would cause these 
galaxies to evolve to nearly axisymmetric shapes after a 
sufficiently long time even in the absence of nuclear black holes.$^{54}$
But the orbital evolution induced by even the steepest cusps
is no more rapid than that caused by a black hole with 
$\rati\approx 0.005$,$^{55}$ the average value for the 
elliptical galaxies in Figure 1.
It follows that black holes are probably more important than 
central density cusps for producing orbital evolution in the 
majority of elliptical galaxies.
To the extent that $\rati$ is independent of $M_g$,
dynamical evolution timescales in triaxial galaxies 
should therefore scale roughly with their crossing 
times, as assumed above.

Kormendy and Bender$^{56}$ have noted that many dynamical 
properties of elliptical galaxies correlate with their isophotal 
shapes, either disky or boxy.
Boxy Es rotate more slowly than disky Es and are more 
likely to exhibit the dynamical signatures of triaxiality.
An interesting question, not addressed by Kormendy and Bender, 
is why triaxiality should correlate with boxiness.
It is tempting to associate ``boxy isophotes'' with ``box 
orbits'' and hence with triaxiality but the tube orbits that make
up axisymmetric galaxies are equally good at generating boxy 
isophotes.$^{57}$
Boxiness results from a non-smooth population of phase space, 
i.e. from a distribution function that is peaked around orbits 
with a narrow range of shapes,$^{58}$ whether 
tubes or boxes.
Eliminating boxiness requires a ``smoothing out'' of
phase space and the orbital evolution induced by a central 
black hole has just this effect.
$N$-body simulations provide some support for this idea:
boxy, triaxial systems evolve -- following an increase 
in their central densities -- into axisymmetric ones with accurately 
elliptical isophotes.$^{35}$
Boxy isophotes may therefore be an indication that the central 
black hole has not had sufficient time (reckoned since the most 
recent merger, say) to strongly influence the orbital distribution,
hence to destroy the triaxiality.
Boxy galaxies constructed primarily from tube orbits would not be
affected by a central mass, which may explain the 
persistence of strong boxiness at the centers of many disk 
galaxies$^{59}$ and in some faint, rapidly-rotating 
ellipticals.$^{60}$

The smoothing effect that a black hole has on the 
orbital population of a triaxial galaxy is an example of 
``chaotic mixing,'' the tendency of the phase space density 
in a chaotic system to approach a constant value at all 
points on the energy surface.$^{61,62}$
Chaotic mixing is responsible for the fact that complex dynamical 
systems like gases exhibit a much narrower range of properties than they 
would if their constituent particles were free to populate phase space in 
arbitrary ways.$^{63}$
In the same manner, the observed regularity in elliptical 
galaxy properties may be attributable in part to chaotic 
mixing induced by nuclear black holes.
One example, just discussed, is the tendency of black holes to 
convert boxy, triaxial galaxies into axisymmetric ones with elliptical 
isophotes -- i.e. to restrict the allowed range of shapes.
On a deeper level, the distribution function of an axisymmetric
galaxy formed in this way would be biased toward forms for 
$f$ that are as nearly constant as possible.
The most general $f$ for an axisymmetric galaxy has the form 
$f(E,L_z,I_3)$ with $I_3$ the third integral.
There are generally many three-integral $f$'s that are consistent 
with a given axisymmetric density law $\rho(\varpi,z)$;$^{64}$ among 
these, the smoothest $f$ is the one that is constant with 
respect to the third integral, $f=f(E,L_z)$.
This argument suggests that the phase space density of stars in 
an axisymmetric galaxy that had evolved, via black-hole-induced 
chaos, from a more general triaxial shape should depend only 
weakly on the third integral.
It is interesting that detailed modelling of M32, a galaxy that 
is old compared to expected timescales for chaotic mixing,$^{62}$ 
yields a best-fit $f\approx f(E,L_z)$.$^{65}$
The extreme non-uniqueness of galactic models,$^{66}$ for so long 
the bane of stellar dynamicists, may largely disappear once the constraints imposed 
by black-hole-induced evolution are more fully understood.

\bigskip

This work was supported by NSF grants AST 93-18617 and AST 
96-17088 and by NASA grant NAG 5-2803.
I thank E. Athanassoula, J. Barnes, F. Combes, D. Friedli, C. Joseph, 
S. McGaugh, A. Quillen, J. Sellwood and M. Valluri for useful 
discussions and for comments on the manuscript.

\end{document}